\documentstyle[preprint,aps,epsfig]{revtex}
\begin{document}

\draft
\preprint{
\vbox{
\hbox{DOE/ER/40762-086}
\hbox{U.MD. PP\# 96-106}
\hbox{LA-UR 96-1960}
}}

\title{Large-$x$ $d/u$ Ratio in $W$-boson Production}

\author{W. Melnitchouk}
\address{Department of Physics,
	 University of Maryland,
 	 College Park, MD 20742}
\author{J.C. Peng}
\address{Physics Division,
	 Los Alamos National Laboratory,
	 Los Alamos, NM 87545}

\maketitle

\begin{abstract} 
Recent analysis of proton and deuteron deep-inelastic scattering data 
have suggested that the extracted $d/u$ quark distribution ratio at 
large $x$ may be significantly larger than previously believed, 
if the data are corrected for nuclear binding effects in the deuteron. 
We examine the sensitivity to the large-$x$ $d/u$ ratio of lepton
asymmetries from $W$-boson production in $p\overline p$ and $pp$ 
collisions at large rapidity, which do not suffer from nuclear 
contamination.
\end{abstract}

\pacs{PACS numbers: 12.38.Qk, 13.38.Be, 13.85.Qk}
										

In recent years our knowledge of quark distributions in the 
nucleon has increased significantly, thanks to the accumulation of 
high-quality deep-inelastic lepton--proton and lepton--deuteron 
scattering data extending over a large range of Bjorken-$x$ and $Q^2$.
Attention has been focussed primarily on unraveling the detailed 
structure of the nucleon sea at $x \alt 0.1$.
The $\overline u - \overline d$ asymmetry \cite{SU2}, for example, 
has been carefully probed by the New Muon and NA51 Collaborations 
at CERN \cite{NMC}, and more recently the CCFR Collaboration \cite{CCFR} 
has investigated the possibility that the strange and antistrange
quark distributions in the nucleon could be different \cite{SSBAR}.

It has usually been taken for granted, on the other hand, that the 
structure of the valence quarks in the nucleon is well known and 
uncontroversial.
Unlike for the sea quarks, charge and baryon number conservation 
fix the first moments of the valence distributions, so that the 
physics of these is contained entirely in the shapes of their 
$x$-distributions.
Nevertheless, the precise $x$-dependence of the $u$ and $d$ quark 
distributions is rather important from the point of view of understanding
spin-flavor symmetry breaking in the proton.
Although the simple spin-flavor SU(2)$\times$SU(2) symmetric expectation 
of $u(x) = 2 d(x)$ is clearly violated at large $x$ \cite{CLOSE79},
the limiting behavior of $d/u$ as $x \rightarrow 1$ is very sensitive 
to the dynamics underlying the symmetry breaking mechanism \cite{NP}.

The softening of the valence $d$ quark distribution relative to the $u$ 
was correlated by Close \cite{CLOSE73} with the $N$--$\Delta$ mass splitting, 
through the observation that both phenomena could be attributed to a larger
mass for the spectator $qq$ pair with spin 1 compared with a scalar $qq$ pair.
Such a mass shift can be produced for example by pion exchange, instantons,
or by a color-magnetic force \cite{COLMAG}.
In the extreme limit of the scalar $qq$ configuration being the only one
relevant at $x=1$, the $d/u$ ratio would be zero.
However, in a model of SU(2)$\times$SU(2) breaking based on a perturbative
treatment of one gluon exchange between the spectator quarks \cite{FJ}, 
the dominant $qq$ configurations at large $x$ were found to be those with 
spin projection zero, so that the interacting quark carries the helicity
of the proton, leading to the prediction $d/u \rightarrow 1/5$.

Experimental information on $d/u$ at large $x$ has usually come
from measurements of the $F_2^n/F_2^p$ structure function ratio,
with the neutron structure function $F_2^n$ extracted from $F_2^p$
and the deuteron $F_2^D$ structure functions \cite{SLAC1,SLAC2,EMC}.
Early analyses suggested that the $d/u$ ratio could be fitted as
$d/u = 0.57 (1-x)$ \cite{EHLQ}.
However, typical analyses of deep-inelastic deuteron data have usually 
been performed under the assumption that the deuteron is a system of 
two free nucleons moving with some Fermi momentum.
While this assumption is perfectly adequate at small and intermediate
values of $x$, at large $x$ ($x \agt 0.7$) the effects of nuclear binding,
in addition to nucleon Fermi motion, can significantly alter the extracted 
$d/u$ ratio \cite{KU,MST}.
In particular, the analysis in Ref.\cite{NP} suggested that inclusion of 
the EMC effect in the deuteron can raise the asymptotic value of $d/u$ 
from the traditional zero result \cite{EHLQ} to a value broadly consistent
with the perturbative QCD expectation of 1/5 \cite{FJ}.

In Fig.1 we illustrate the sensitivity of the $d/u$ ratio to the 
treatment of the nuclear effects in the deuteron.
The full (open) circles are derived from the SLAC data \cite{SLAC1,SLAC2}
assuming Fermi motion plus binding (Fermi motion only) corrections 
\cite{NP}.
The dashed curves represent the valence $d/u$ ratio as given by 
standard parameterizations of global parton distributions in 
Ref.\cite{CTEQ} (the results for other parameterizations \cite{MRS,GRV}
are in fact very similar).
Because the large-$x$ $d/u$ ratio in the standard parameterizations 
\cite{CTEQ,MRS,GRV} is fitted to the proton and ``neutron'' structure 
function data, obtained from $F_2^p$ and $F_2^D$ without inclusion 
of nuclear EMC corrections in the deuteron \cite{SLAC1,EMC}, the $d/u$
ratio is seen to approach zero as $x \to 1$.
To allow for the possibility that the $x \rightarrow 1$ limit of $d/u$ 
is non-zero, we modify the $d$ quark distribution according to:
\begin{eqnarray}
\label{dmod}
{d(x,Q^2) \over u(x,Q^2)}
&\rightarrow& {d(x,Q^2) \over u(x,Q^2)} + \Delta(x,Q^2).
\end{eqnarray}
The analysis of Ref.\cite{NP} suggested that at $Q^2 \approx 10$ GeV$^2$
the correction term could be parameterized by the simple form:
\begin{eqnarray}
\label{Delta10}
\Delta(x,Q^2=10{\rm GeV}^2) 
&\approx& 0.2\ x^2\ \exp(-(1-x)^2),
\end{eqnarray}
so that $\Delta \rightarrow 1/5$ in the limit $x \rightarrow 1$ \cite{FJ}.
(The overall normalization of the valence $d$ quark distribution can be 
preserved by a correspondingly small change at small $x$, however, such 
a change will hardly be noticeable in the final cross section ratio.)
The result with the modified $d$ quark distribution is shown by the 
solid curves in Fig.1.
The effect of evolution on the $d/u$ ratio, though sizable at low $x$ 
values, is quite small at large $x$, as seen by evolving \cite{WM}
the curves in Fig.1 from $Q^2=10$ GeV$^2$ to $Q^2=M_W^2$.

Given the sensitivity of the $d/u$ ratio obtained from the proton 
and deuteron data to theoretical inputs for the deuteron structure, 
to resolve the issue one would obviously like to appeal to data 
without the need to model the nuclear effects in the deuteron.
One such possibility is neutrino and antineutrino scattering from
the proton.
Because the $\nu p$ structure function at large $x$ is determined
essentially by the $d$ quark distribution, while the $\overline \nu p$
by the $u$ distribution, combining the two sets allows one to obtain 
a model-independent determination of $F_2^n/F_2^p$ in the valence 
quark dominated region.
Unfortunately, the existing neutrino data \cite{NU} do not extend 
to very large $x$ ($x \alt 0.6$), and the error bars are too large 
to discriminate between the different $d/u$ behaviors.
Yet another possibility to reduce the sensitivity to nuclear
effects is to perform semi-inclusive measurements on the deuteron,
tagging a proton in the final state \cite{SEMI}.
If the momentum of the produced proton is small, one can probe nearly
on-shell neutrons, although the required energy in such experiments
must be sufficiently large to obtain a clean separation of the current
and target fragmentation regions.
final state interactions.


In this paper we investigate the possibility of pinning down the large-$x$ 
$d/u$ ratio through $W$-boson production in $p \overline p$ and $pp$ 
collisions.
As noted by Berger et al. \cite{BHKW} (see also \cite{MRSW}), this method
has the virtue that it is completely free of the theoretical ambiguities
associated with modeling the nuclear physics in the deuteron
\cite{NP,SLAC1,SLAC2,KU,SEMI}.
An additional advantage, which has not been discussed previously, is that
one does not need to rely on any assumptions about charge symmetry in the
nucleon.
Some recent model calculations \cite{CSB} of quark distributions in the
proton and neutron have suggested potentially non-negligible charge symmetry
breaking effects, which would further contaminate the $d/u$ ratio extracted
from $H$ and $D$ data.

Because $W^+$ and $W^-$ bosons couple to different quark flavors in the 
proton, any differences between $u$ and $d$ quark distributions will be
reflected through different $W^\pm$ rapidity distributions.
This property has previously been suggested as a means of constraining
the $\overline d/\overline u$ ratio at small $x$ \cite{BHKW,MRSW,ES,PJ}.
In fact the CDF Collaboration at Fermilab \cite{CDF95,BODEK} has measured
the lepton charge asymmetry resulting from the decay $W \rightarrow l \nu$ 
(where $l = e, \mu$), although in the rapidity range considered, these data
were sensitive mostly to the quark distributions at $x \alt 0.3$, which
unfortunately is too low to discriminate between the different large-$x$ 
valence $d/u$ behaviors.
By considering charge asymmetries at large lepton rapidities, corresponding
to highly energetic decay leptons, we will show that one can probe the
valence $u$ and $d$ distributions also in the large-$x$ region where
SU(2)$\times$SU(2) symmetry breaking effects are dominant.

At leading order, the differential cross section for the production of 
a $W$ boson in a proton--$h$ collision (where the hadron $h=\overline p$ 
or $p$) is proportional to \cite{BP}:
\begin{mathletters}
\label{sigW}
\begin{eqnarray}
\label{sigW+}
{d\sigma \over d x_F} (W^+)
&=& K {2\pi G_F \over 3 \sqrt{2}} 
\left( {x_1 x_2 \over x_1 + x_2} \right)
\left\{ \cos^2\theta_c\ 
        \left(  u_p(x_1)\ \overline{d}_h(x_2)\  
     +\ \overline{d}_p(x_1)\ u_h(x_2)
    	\right)\ 
\right.		\nonumber\\
& & \hspace*{3cm}
\left.
\ +\    \sin^2\theta_c\
	\left(  u_p(x_1)\ \overline{s}_h(x_2)\ 
      	     +\ \overline{s}_p(x_1)\ u_h(x_2)
	\right) 
\right\}, 
\end{eqnarray}
where $x_F \equiv x_1 - x_2$.
With this definition of $x_1$ and $x_2$, positive $x_F$ is defined 
along the direction of the proton.
Furthermore, $x_1 x_2 = M_W^2/s$, where $s$ is the total $p$--$h$ 
center of mass energy squared, and $M_W$ is the $W$-boson mass. 
(The variable $x_F$ is related to the $W$-boson rapidity $y$ by 
$x_F = (M_W/\sqrt{s}) (e^y - e^{-y})$).  
In Eq.(\ref{sigW+}) $G_F$ is the Fermi coupling constant, 
$\theta_c$ the Cabibbo angle, and the factor $K$ contains
QCD radiative corrections.
Similarly one can write the $W^-$ differential cross section as:
\begin{eqnarray}
\label{sigW-}
{d\sigma \over d x_F} (W^-)
&=& K {2\pi G_F \over 3\sqrt{2}}
\left( {x_1 x_2 \over x_1 + x_2} \right)
\left\{ \cos^2\theta_c\ 
        \left(  d_p(x_1)\ \overline{u}_h(x_2)\
             +\ \overline{u}_p(x_1)\ d_h(x_2)
        \right)\
\right.         \nonumber\\
& & \hspace*{3cm}
\left.
\ +\    \sin^2\theta_c\
        \left(  s_p(x_1)\ \overline{u}_h(x_2)\
             +\ \overline{u}_p(x_1)\ s_h(x_2)
	\right)
\right\}.
\end{eqnarray}
\end{mathletters}%
Taking the ratio of the $W^+$ and $W^-$ cross sections for $p$--$h$ 
scattering (c.f. the ratio constructed by Berger et al. \cite{BHKW}
in terms of an asymmetry between $W^+$ and $W^-$ cross sections),
one can then define:
\begin{eqnarray}
R_{p h}(x_F)
&\equiv& {d\sigma/dx_F\ (W^+) \over d\sigma/dx_F\ (W^-)}.
\end{eqnarray}

Production of $W$-bosons in $p \overline p$ collisions has been studied 
by the CDF Collaboration \cite{CDF95,BODEK,CDF92} at the Tevatron, where
the total center of mass energy is $\sqrt{s} = 1.8$ TeV.
One can see how $R_{p \overline p}$ is related to the $d/u$ ratio by
neglecting the small sea-sea and $\sin^2\theta_c$ terms in Eqs.(\ref{sigW}),
in which case:
\begin{eqnarray}
R_{p \overline p}(x_F) 
&\approx& {u(x_1) \over d(x_1)} \cdot {d(x_2) \over u(x_2)}, 
\end{eqnarray}
where we have used $\overline q_{\overline p}(x) = q_p(x)$ 
to relate all distributions to those in the proton.
To date the CDF Collaboration, and most theoretical interest, has focused
on constraining quark densities in the small-$x$ region ($x \alt 0.3$).
To probe the $u/d$ ratio in the proton at large $x_1$ requires large 
values of $x_F$ (or equivalently large $W$-boson rapidities $y$).
For $x_1 \agt 0.5$, a center of mass energy $\sqrt{s}=1.8$ TeV implies 
$x_2 \alt 0.004$, so that $x_F \approx x_1$.

In order to demonstrate the sensitivity of the ratio of the $W^+$ to $W^-$ 
cross sections to $d/u$ at large $x$, in Fig.2(a) we compare $R_{p\overline p}$
for the two different parameterizations of the $d$-quark distribution used
in Fig.1. 
Evolving the $u$ and the modified and unmodified $d$ quark distributions
to the $W$-boson scale, one can extract from Eq.(\ref{dmod}) the correction
$\Delta(x,Q^2)$ at $Q^2 = M_W^2$, which to a good approximation can be
parameterized as:
\begin{eqnarray}
\label{DeltaMW2}
\Delta(x,Q^2=M_W^2)
&\approx& 0.18\ x^{1.2}\ \exp\left(-(1-x)^2\right).
\end{eqnarray}
The ratio $R_{p\overline p}$ with the modifed and unmodified $d$ quark 
distributions is shown by the solid and dashed curves in Fig.2(a) (in the 
numerical calculations the full expressions from Eqs.(\ref{sigW}) are used).
As can be seen, the modified distribution already introduces a reduction 
in the ratio of $\sim 30\%$ at $x_F = 0.6$, and more than 50\% for 
$x_F \agt 0.7$.
Indeed, if one allows $d/u \rightarrow 0$, the ratio $R_{p\overline p}$,
which depends on $u(x_1)/d(x_1)$, increases quite dramatically as
$x \rightarrow 1$.

Of course measurement of this ratio at these rather large values of $x_F$
will be difficult in view of the falling cross sections there.
However, by increasing the luminosity and rapidity coverage anticipated 
in future runs at CDF \cite{AB}, one should be able to access this region
with reasonable statistics.
In addition, the statistics can be improved by a factor of two by 
combining data from $x_F > 0$ with those from $x_F < 0$, since the 
ratio is inversely symmetric about $x_F=0$, 
i.e. $R_{p\overline p}(x_F) = 1/R_{p\overline p}(-x_F)$.

The large-$x$ $d/u$ ratio can also be probed in $pp$ collisions,
for example in future at RHIC.
This is again evident by dropping the Cabibbo angle suppressed 
terms in Eqs.(\ref{sigW}) (with $h=p$), so that the ratio $R_{pp}$
of the $W^+/W^-$ cross sections can be approximated by:
\begin{eqnarray}
R_{pp}(x_F)
&\approx& {u(x_1)\ \overline d(x_2) + \overline d(x_1)\ u(x_2) 
     \over \overline u(x_1)\ d(x_2) + d(x_1)\ \overline u(x_2)}.
\end{eqnarray}
Doncheski et al. \cite{DHKS} have recently studied $W$-boson asymmetries
at RHIC energies as a means of determining $\overline d - \overline u$
differences in connection with the Gottfried sum rule, as well as
$\Delta u - \Delta d$ differences in polarized parton distributions.
As with the earlier $p\overline p$ studies at CDF, however, attention has
been focussed exclusively on the sea quark distributions in the small-$x$
region.
In contrast, here we point out that RHIC $pp$ collisions can also be used
to obtain information on the valence quark distributions at large $x$.
It was also suggested in Ref.\cite{NAD} that one could measure the ratio
$\Delta d/d$ at large $x$ in polarized $p\vec{p}$ collisions at RHIC,
however, in view of the present discussion it is clear that the unpolarized
$d$ quark distribution itself needs to be accurately determined before any 
conclusions can be drawn about $\Delta d$.

For a RHIC center of mass energy $\sqrt{s} = 500$ GeV, for $x_1 \agt 0.6$
at large $x_F$ one will have $x_2 \alt 0.04 \ll x_1$, so that one may expect
for the antiquark distributions
$\overline u(x_1) \ll \overline u(x_2)$ and 
$\overline d(x_1) \ll \overline d(x_2)$, 
in which case $R_{pp}$ will directly probe the valence $u/d$ ratio:
\begin{eqnarray}
R_{pp}(x_F)
&\approx& {u(x_1) \over d(x_1)} 
    \cdot {\overline d(x_2) \over \overline u(x_2)},\ \ \ \ x_1 \gg x_2.
\end{eqnarray}
In Fig.2(b) we compare the ratio $R_{pp}(x_F)$ for the standard CTEQ
parameterization \cite{CTEQ} (dotted curve) with that calculated with
the modified $d$ distribution in Eq.(\ref{dmod}) (solid curve).
As for the $p\overline p$ case, the latter is significantly smaller
for $x_F \agt 0.5$ than the ratio with the standard parameterization.
We emphasize that these effects are considerably larger than those
found in earlier analyses \cite{BHKW} by comparing standard parton
parameterizations.

Although the differences between the $x_F$ ratios with the unmodified
and modified distributions for both $p\overline p$ and $pp$ collisions
are quite conspicuous at large $x_F$, in practice it is not the 
$x_F$ distributions which are measured but rather the charge asymmetry 
of the leptons resulting from the $V$--$A$ decay of the $W$ bosons.
The measured lepton asymmetry is defined by \cite{MRS90}:
\begin{eqnarray}
A(y_l) 
&=& { d\sigma/dy_l (l^+)\ -\ d\sigma/dy_l (l^-)
\over d\sigma/dy_l (l^+)\ +\ d\sigma/dy_l (l^-)},
\end{eqnarray}
where the lepton rapidity 
$y_l = 1/2 \ln \left[(E_l + p_l)/(E_l - p_l)\right]$
is defined in terms of the decay lepton's energy $E_l$ and longitudinal
momentum $p_l$ in the laboratory frame.
The differential cross section $d\sigma/dy_l$ is obtained by convoluting 
the $q\overline q \rightarrow W$ cross section for each $x_F$ with the 
relevant $W \rightarrow l\ \nu$ decay distribution,
$d\sigma/d\cos\theta \propto (1 \pm \cos\theta)^2$, 
where $\theta$ is the angle between the lepton $l^{\pm}$ direction
and the $W^{\pm}$ polarization in the $W$ rest frame.

In Fig.3 we show the lepton asymmetry $A(y_l)$ calculated for the 
standard CTEQ parton parameterization \cite{CTEQ} (dotted) and with 
the modified $d$ quark distribution, Eq.(\ref{dmod}), (solid).
The asymmetry for $y_l < 0$ is equal and opposite to that for positive
$y_l$.
To compare with the CDF lepton asymmetry data \cite{CDF95,BODEK} we
have made a cut in the transverse energy of the lepton of $E_T > 25$ GeV
in Fig.3(a).
With this cut the $x_F$ distributions at large $x_F$, which are relatively
small in magnitude, are overwhelmed by the much larger cross sections at
smaller $x_F$, where the unmodified and modified distributions are 
essentially equivalent.
In order to resolve the differences at large $x_F$ in the $y_l$ 
distributions one should select from the lepton spectrum only
the most highly energetic events.
In practice, for the CDF center of mass energy $\sqrt{s} = 1.8$ TeV,
this requires a tighter cut of $E_T \agt 35$ GeV on the transverse 
lepton energy.
Such a cut favors the large-$x_F$ distributions, so that for large lepton 
rapidities ($y_l \agt 2$) the differences between valence quark distributions
at large $x_1$ will leave quite noticeable traces, Fig.3(b).
Note that the latest data \cite{BODEK} already extend up to $y_l \approx 2.2$,
with quite small error bars even at the largest rapidities.
With an upgraded Tevatron, a rapidity coverage of at least up to $y_l = 2.5$
is expected \cite{AB}, where the effects are of the order of 50\%.
Together with a luminosity upgrade, this should allow a significant
improvement in the statistics so as to pose a serious test for the
two scenarios for $d/u$ at large $x$.


In summary, $W$-boson asymmetries in $pp$ and $p\overline p$ collisions 
can be used to constrain the behavior of valence quark distributions in 
the large-$x$ region.
The principle advantage is that one can avoid the theoretical ambiguities 
associated with the traditional method of extracting quark distributions
from proton and deuteron deep-inelastic scattering data, which inevitably
requires modeling the nuclear effects in the deuteron.
%
%
We have investigated the sensitivity of the asymmetries to different 
asymptotic behaviors of the valence $d/u$ ratio, and find the ratio
of $W^+/W^-$ cross sections to be very sensitive to the precise
$x \rightarrow 1$ dependence of $d/u$ for values of $x_F \agt 0.5$.
By measuring of the resulting lepton charge asymmetry at large lepton
rapidities one can determine this ratio experimentally by selecting
the more energetic decay leptons.
The availability of such data from an upgraded Tevatron or RHIC would
be instrumental in pinning down the asymptotic behavior of $d/u$,
and therefore should help unravel the physics behind the mechanism of
flavor symmetry breaking in the nucleon.

\acknowledgements

We wish to thank S.J. Brodsky, F.E. Close, T. Goldman, W.-Y.P. Hwang, 
X. Ji, J.M. Moss, and A.W. Thomas for helpful discussions.
We thank the Institute for Nuclear Theory at the University of Washington
for its hospitality and support during recent visits, where this work was
conceived. 
This work was supported by the Department of Energy grant 
DE-FG02-93ER-40762.



\begin{figure}
\epsfig{figure=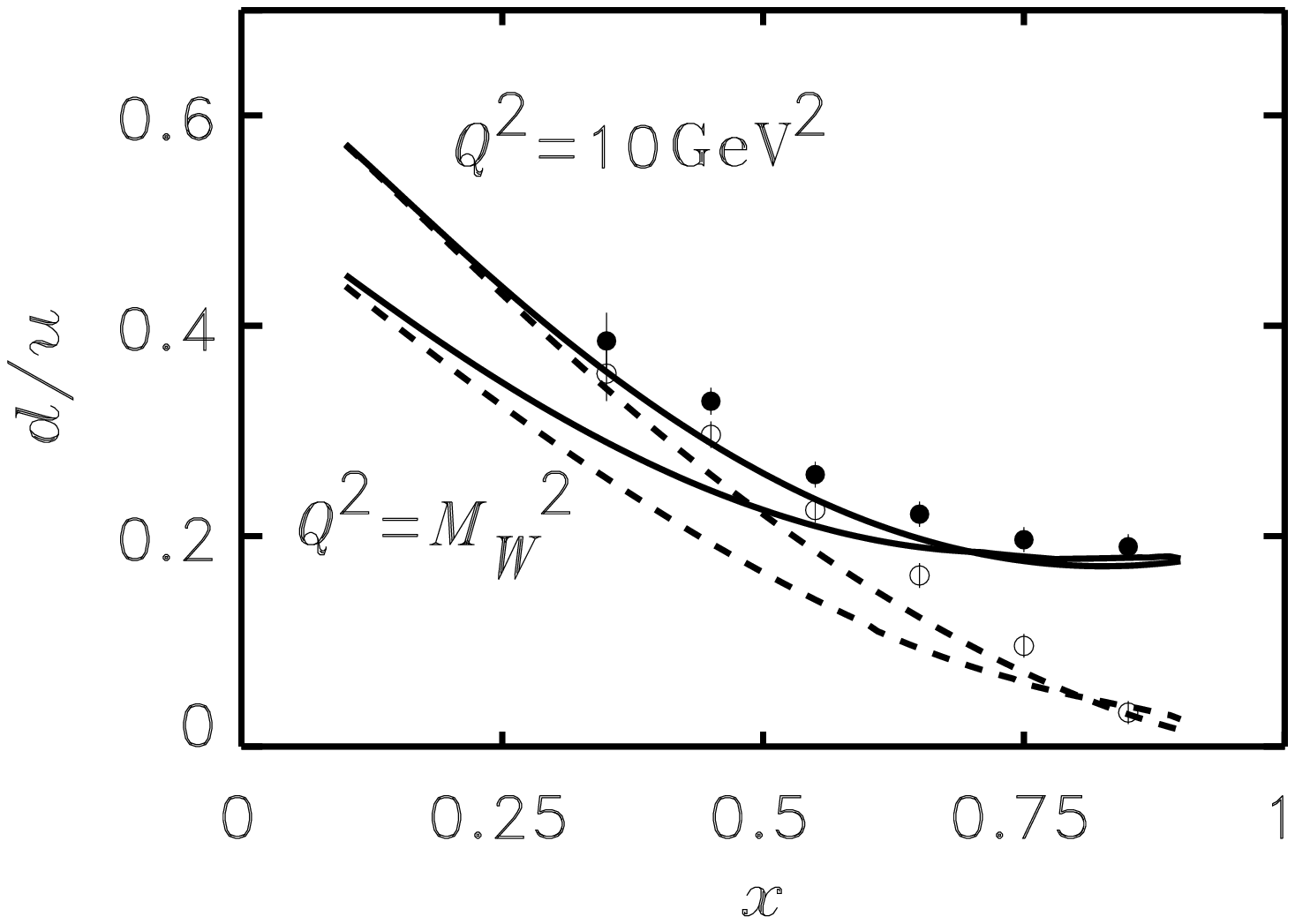,height=9cm}
\caption{$d/u$ quark distribution ratio at $Q^2=10$ GeV$^2$ and 
	$Q^2=M_W^2$.  Dashed curves are parameterizations from 
	Ref.\protect\cite{CTEQ}, solid curves include the modified
	$d$ quark distribution in Eqs.(\protect\ref{dmod}), 
	(\protect\ref{Delta10}) and (\protect\ref{DeltaMW2}).
	Full (open) circles represent SLAC data from 
	Refs.\protect\cite{SLAC1,SLAC2}
	analyzed assuming binding plus Fermi motion (Fermi motion only)
	corrections in the deuteron \protect\cite{NP}.}
\end{figure}

\begin{figure}
\epsfig{figure=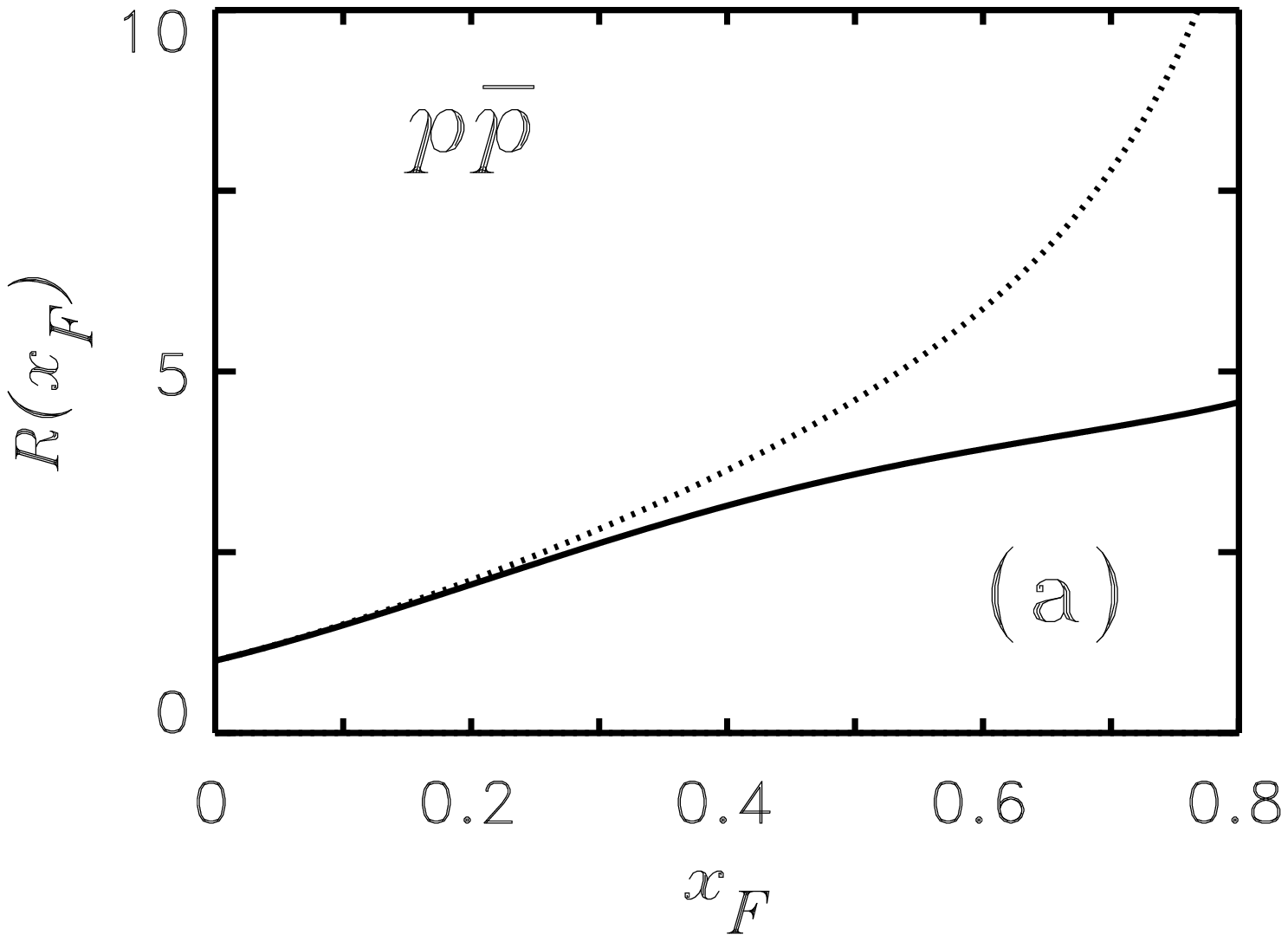,height=9cm}
\epsfig{figure=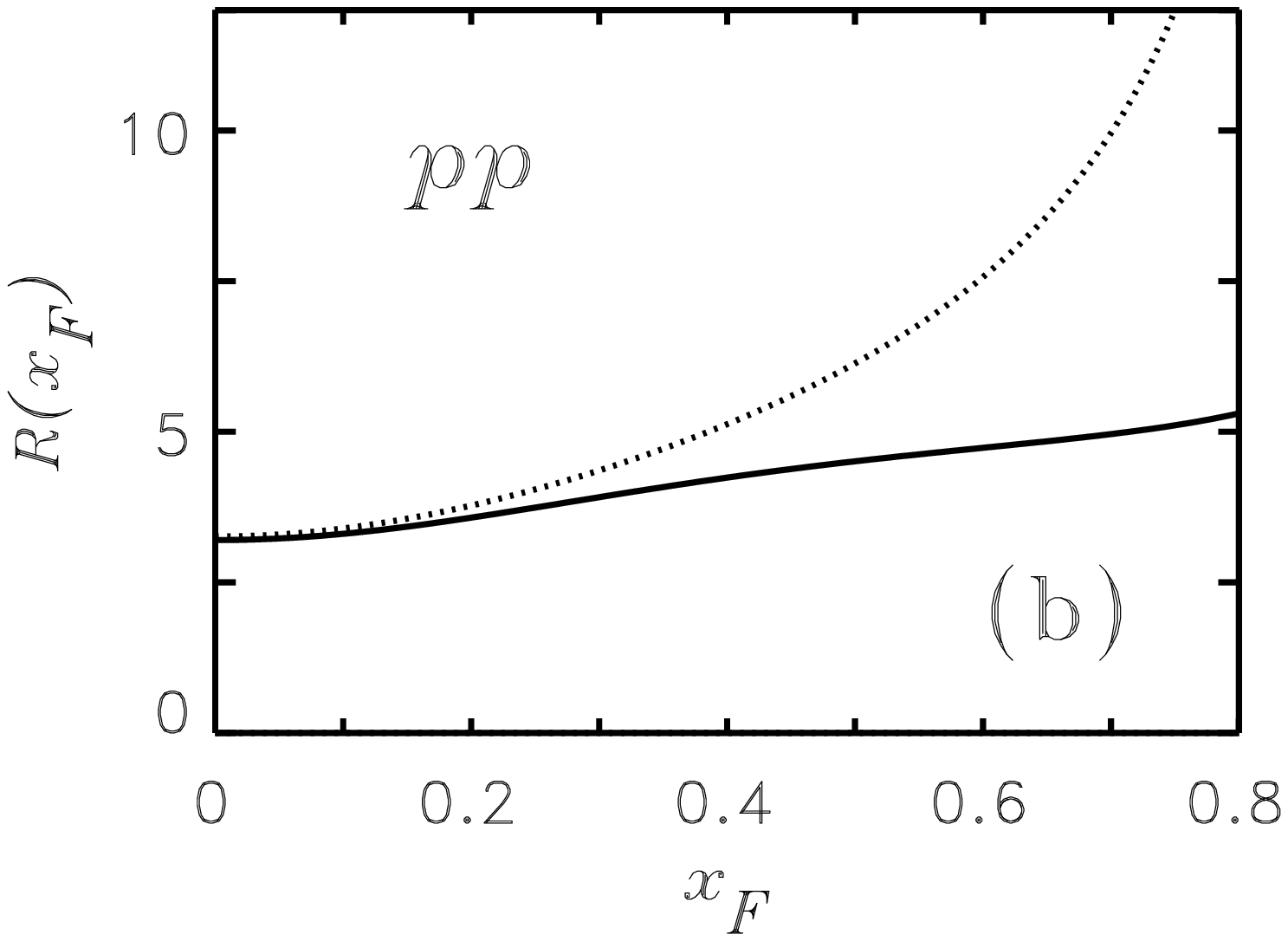,height=9cm}
\caption{Ratio of $W^+/W^-$ cross sections as a function of $x_F$ in 
	(a) $p\overline p$ collisions at a center of mass energy 
	$\protect\sqrt{s} = 1.8$ TeV, and 
	(b) $p p$ collisions at $\protect\sqrt{s} = 500$ GeV.
	The results for the standard CTEQ parton parameterization
	\protect\cite{CTEQ} are represented by the dotted curves, 
	while the solid are for the $d$ quark distribution modified
	according to Eq.(\protect\ref{dmod}).}
\end{figure}

\begin{figure}
\epsfig{figure=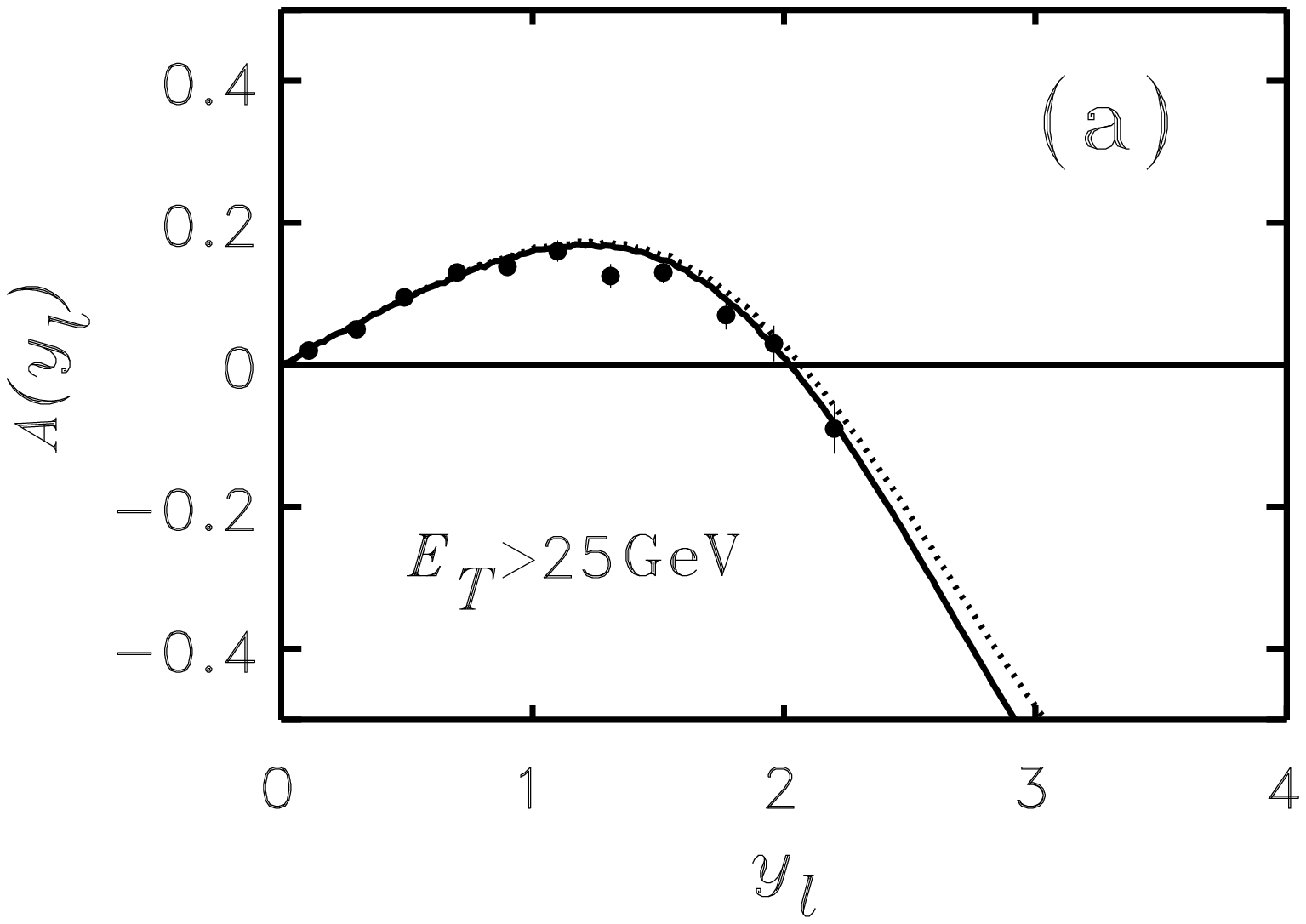,height=9cm}
\epsfig{figure=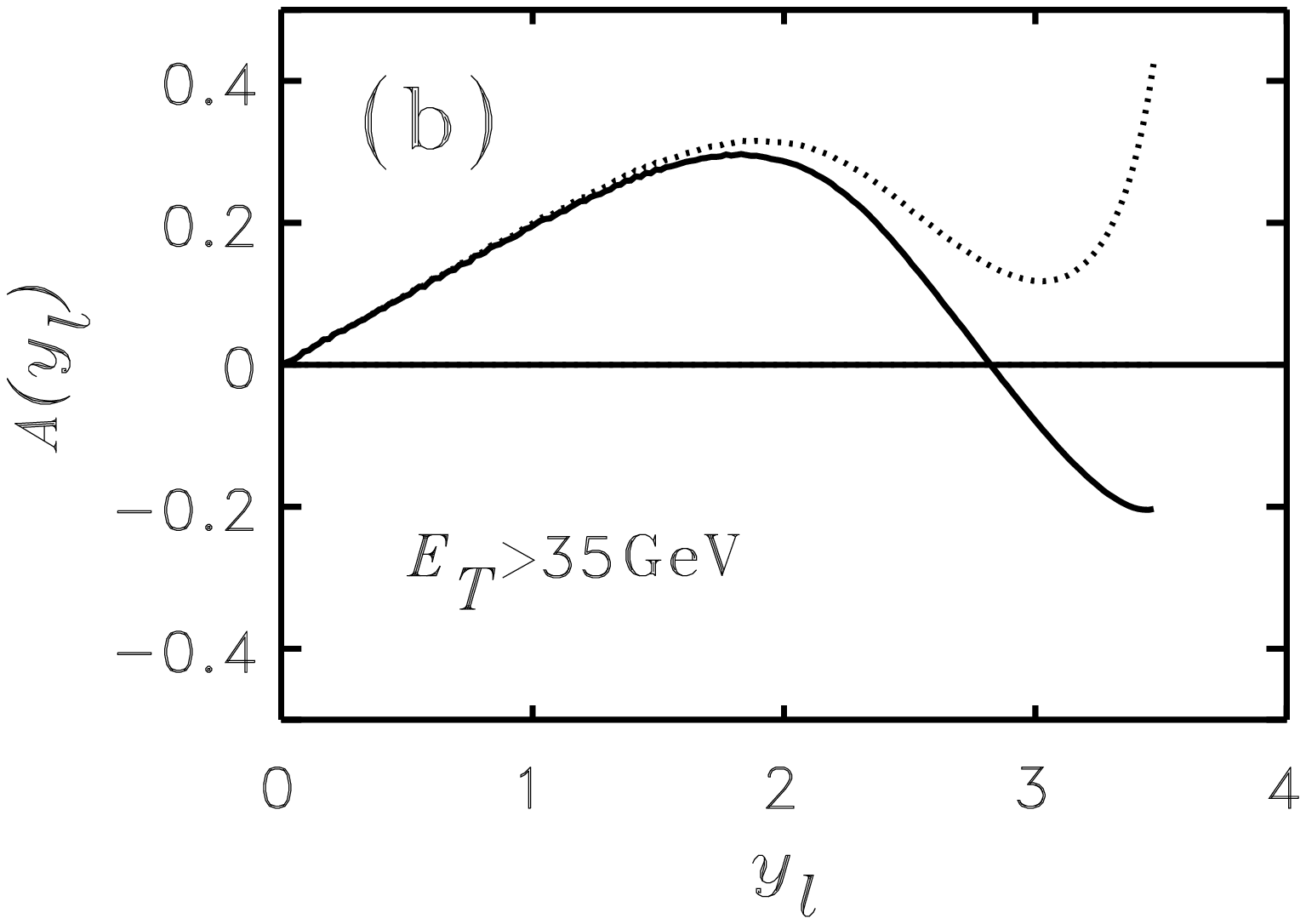,height=9cm}
\caption{Lepton charge asymmetry as a function of lepton rapidity $y_l$
	in $p \overline p$ collisions for $\protect\sqrt{s} = 1.8$ TeV.
	(a) $E_T > 25$ GeV cut for comparison with data from CDF 
	\protect\cite{CDF95,BODEK};
	(b) $E_T > 35$ GeV cut.
	Solid and dotted curves are as in Fig.2.}
\end{figure}

\end{document}